\documentclass[epjST]{svjour}
\usepackage{graphics}

\usepackage{eucal}
\usepackage{amssymb}

\newcommand{\be}{\begin{eqnarray}}
\newcommand{\ee}{\end{eqnarray}}
\newcommand{\bse}{\begin{subequations}}
\newcommand{\ese}{\end{subequations}}

\newcommand{\bnum}{\begin{enumerate}}
\newcommand{\enum}{\end{enumerate}}

\newcommand{\bit}{\begin{itemize}}
\newcommand{\eit}{\end{itemize}}

\newcommand{\bc}{\begin{cases}}
\newcommand{\ec}{\end{cases}}

\newcommand{\bpm}{\begin{pmatrix}}
\newcommand{\epm}{\end{pmatrix}}

\newcommand{\bvm}{\begin{vmatrix}}
\newcommand{\evm}{\end{vmatrix}}

\newcommand{\bs}{\vec}

\newcommand{\mcal}{\mathcal}

\newcommand{\mrm}{\mathrm}

\newcommand{\ovl}{\overline}

\newcommand{\ga}{\alpha}
\newcommand{\gb}{\beta}
\newcommand{\gc}{\gamma}
\newcommand{\gd}{\delta}
\newcommand{\gl}{\lambda}

\newcommand{\go}{\omega}

\newcommand{\gr}{\rho}

\newcommand{\gs}{\sigma}

\newcommand{\Gd}{\Delta}

\newcommand{\Gl}{\Lambda}

\newcommand{\gvf}{\varphi}
\newcommand{\gvr}{\varrho}

\newcommand{\p}{\partial}
\newcommand{\f}{\frac}
\newcommand{\diff}{\mrm{d}}

\newcommand{\Hyd}{\mcal{H}}

\newcommand{\csp}{\;,\qquad}

\newcommand{\dX}{\diff X}

\newcommand{\dt}{\diff t}
\newcommand{\ds}{\diff s}

\newcommand{\met}{\mrm{m}}

\renewcommand{\sec}{\mrm{s}}

\newcommand{\kg}{\mrm{kg}}

\begin{document}

\title{Low Reynolds number hydrodynamics of asymmetric, oscillating dumbbell pairs}
\author{
Victor B. Putz\inst{1}\fnmsep\thanks{\email{v.putz1@physics.ox.ac.uk}} 
\and 
J\"orn Dunkel
\inst{1} 
\fnmsep\thanks{\email{j.dunkel@physics.ox.ac.uk}} 
}
\institute{Rudolf Peierls Centre for Theoretical Physics, University of Oxford, 1 Keble Road, Oxford OX1 3NP, United Kingdom}
% \and the second here \and ...}
%
\abstract{
Active dumbbell suspensions constitute one of the simplest model system for collective swimming at low Reynolds number. Generalizing recent work, we derive and analyze stroke-averaged equations of motion that capture the  effective hydrodynamic far-field interaction between two oscillating, asymmetric dumbbells in three space dimensions.  Time-averaged equations of motion, as those presented in this paper, not only yield a considerable speed-up in numerical simulations,  they may also serve as a starting point when deriving continuum equations for the macroscopic dynamics of multi-swimmer suspensions. The specific model discussed here appears to be particularly useful in this context, since it allows one to investigate how the  collective macroscopic behavior is affected by changes in the microscopic symmetry of individual swimmers.  
} 
%end of abstract
\maketitle

\section{Introduction}
\label{intro}

The rich collective behavior of active biological systems, such as flocks of birds, schools of fish or bacterial suspensions, has attracted considerable interest in biophysics in recent years~\cite{2005ToYuRa,2008EbSG,2008UnHOGr,2009StEtAl,2009RoCoSG,2009BaMa_PNAS}. Due to the complex multi-scale nature of the observable patterns~\cite{2009CoWe,2007SoEtAl},  theoretical approaches are manifold, ranging from discrete \lq microscopic\rq\space descriptions, that account for the dynamics of individual organisms~\cite{2008EbSG,2009StEtAl,2009RoCoSG}, to field-theoretic  continuums models, that aim to capture dynamical features and symmetries on the macro-scale~\cite{2002Ra,2008BaMa,2009BaMa_PNAS}. The arguably most important problem in this context is to understand which microscopic properties determine (or are reflected in) the macroscopic behavior~\cite{2005ToYuRa,2009BaMa_PNAS}.  
\par
Generally, linking micro- and macro-behavior is mathematically difficult as it typically involves both spatial and/temporal averaging (coarse-graining) of nonlinear equations of motion~\cite{2008LaBa,2008AlYe,2009DuZa}. Valuable insight can be gained by considering simplified model systems that allow for explicit testing of averaging procedures~\cite{2010PuDuYe,2010DuPuZaYe}. In the present paper, we intend to study the validity of temporal coarse-graining for a simple model~\cite{2008LaBa,2010PuDuYe} of collective swimming at low Reynolds number, a regime relevant to the motion of bacteria and small algae~\cite{1977Pu,1987ShWi,2009PoEtAl_Gold}. Specifically, generalizing recent work~\cite{2008AlYe,2008LaBa,2010PuDuYe}, we derive time-averaged equations of motion that govern the effective, \emph{three-dimensional} hydrodynamic interaction between actively oscillating, \emph{asymmetric} dumbbell\footnote{A dumbbell is defined as a pair  of rigid spheres (Stokeslets) connected by an oscillating, harmonic spring~\cite{2008LaBa,2010PuDuYe}.} pairs.
\par 
It has been known for a long time~\cite{1977Pu} that --  owing to the time-reversibility of the Stokes-equation -- isolated, force-free dumbbells cannot swim in the zero-Reynolds number limit. However, when the phases of two identical, periodically oscillating dumbbells at finite separation are detuned, then collective swimming by means of hydrodynamic interaction becomes  possible~\cite{2008LaBa}. Thus, dumbbell \lq swimmers\rq\space may be regarded as the simplest  model for collective swimming at low Reynolds number. Related recent work focusses either on the one-dimensional case~\cite{2008LaBa} or on symmetric two-sphere swimmers~\cite{2010PuDuYe}. Here, we generalize the analysis of Ref.~\cite{2010PuDuYe} to the asymmetric case.  The motivation for this is as follows: 
\par
Symmetric dumbbells do not possess an intrinsic orientation. Consequently, a corresponding field-theoretic continuum description must be based on the $Q$-tensor\footnote{$Q$-tensor := second moment tensor of orientations~\cite{2002Ra}}, since their local mean orientation (polarization) field is trivially zero, $\bs \Pi\equiv \vec 0$. By contrast, asymmetric dumbbells may possess a nontrivial mean orientation field, $\bs \Pi(t,\bs x)\ne \vec 0$, which in $d>1$ space dimensions can change due to hydrodynamic interactions. Hence, asymmetric dumbbells appear to be the simplest model system for studying how the symmetry of the microscopic constituents affects both hydrodynamics and orientational order in the continuum limit. 
\par
However, before one can derive the corresponding field equations by means of standard methods~\cite{2009BaMa_PNAS}, one  first has to identify the effective equations of motion for a dumbbell's position and orientation change by averaging the hydrodynamic interactions with the other swimmers over a swimming stroke. It is, therefore,  the purpose of the present paper to (i) provide explicit expressions for the time-averaged, effective interaction forces, and to (ii) verify their validity by comparing the exact microscopic with the coarse-grained dynamics.

%%%%%%%%%%%%%%%%%%%%%%%%
\section{Stroke-averaged equations of motion}
\label{sec:1}
%%%%%%%%%%%%%%%%%%%%%%%%
We first summarize the microscopic equations of motions of the dumbbell model. Subsequently, 
the corresponding stroke-averaged equations will be discussed.

%%%%%%%%%%%%%%%%%%%%%%%%
\subsection{Microscopic model}
\label{s:micro}
%%%%%%%%%%%%%%%%%%%%%%%%

We consider a system of $S$ identical dumbbells. Each dumbbell swimmer consists of two spheres, with radii $a_1$ and $a_2$. At low Reynolds numbers, inertia is negligible and the state of the system at time $t$ is completely described by the spheres' position coordinates $\{\bs X_\ga\}=\{X_{(\ga i)}(t)\}$ with $\ga=1,\ldots, 2S$ labeling the spheres,  and $i=1,2,3$ the space dimension (throughout, we adopt the Einstein summation convention for repeated \emph{Latin} indices).   Neglecting rotations of the spheres, their dynamics is governed by the overdamped equations~\cite{2008LaBa,2009DuZa,2010PuDuYe} 
\be\label{e:langevin}
\dot X_{(\ga i)}(t)
=
\sum_{\gb=1}^{2S}\Hyd_{(\ga i)(\gb j)} F_{(\gb j)}
\label{e:langevin-a}
\ee
where $\dot X:=\dX/\dt$ is the velocity\footnote{Ref.~\cite{2010PuDuYe} discusses how to include thermal fluctuations in Eq.~(\ref{e:langevin}).} The hydrodynamic interaction tensor $\Hyd$ couples the deterministic  force components $F_{(\gb i)}$ that act on the individual spheres. Generally, the vector $F=\{F_{(\gb i)}\}$ may comprise contributions from internal forces, i.e., those required to bind two spheres to form a dumbbell,  as well as from external force fields (gravity, etc.); however, in the present paper, we shall assume that external forces are negligible. 
\par
In our numerical simulations, $\Hyd$ is given by the Rotne-Prager-Yamakawa-Mazur tensor~\cite{1969RoPr,1970Ya,Oseen,HappelBrenner,1982Ma}
\be\label{e:Mazur}
\label{e:Mazur_diagonal}
\Hyd_{(\ga i)(\ga j)}
&=&
\f{\gd_{ ij}}{\gc_{\ga}}=
\f{\gd_{ ij}}{6\pi \mu a_\ga}
\\
\Hyd_{(\ga i)(\gb j)}
&=&
\f{1}{8\pi\mu\, r_{\ga\gb}} 
\biggl(  \gd_{ij} + 
\f{r_{\ga\gb i} r_{\ga\gb j}}{r_{\ga\gb}^2} 
\biggr)+
\label{e:Oseen}
\f{a_\ga^2+a_\gb^2}{24\pi \mu\; r_{\ga\gb}^3}
\biggl(
\gd_{ij} -
3 \f{r_{\ga\gb i} r_{\ga \gb j}}{r_{\ga\gb}^2} 
\biggr),
\qquad
\label{e:Mazur_offdiagonal}
\ee
where $r_{\ga\gb i}:=x_{\ga i}- x_{\gb i}$, $\ga\ne\gb$,  and $r_{\ga\gb}:=|\bs x_\ga-\bs x_\gb|$. Analytical formulas presented below are based on an Oseen approximation, which neglects the 
$r_{\ga\gb}^{-3}$ term in Eq.~(\ref{e:Mazur_offdiagonal}). 
The diagonal components~(\ref{e:Mazur_diagonal}) describe Stokesian friction in a fluid of viscosity $\mu$. The off-diagonal components~(\ref{e:Mazur_offdiagonal})  model hydrodynamic interactions between different spheres.  Note that $\Hyd$ is positive definite for $r_{\ga\gb}>a_\ga+a_\gb$ and divergence-free,  $\sum_{\gb}\p_{(\gb j)} \Hyd_{(\ga i)(\gb j)}\equiv 0$ with $\p_{(\gb i)}:=\p/\p x_{(\gb i)}$.
\par
We still need to specify the intra-dumbbell force $F$. Let us consider the dumbbell~$\gs$, formed by spheres $\ga=2\gs-1$ and $\gb=2\gs$, and denote its length by $d^{\gs}(t):=|\bs X_\gb(t)-\bs X_\ga(t)|$. Neglecting external force fields, we shall assume that the two spheres are connected by a harmonic spring of variable length \be
L^\gs(t)=\ell+\gl \sin(\go t +\gvf^\gs),
\qquad
\qquad
\ell > a_\ga+a_{\gb} +\gl.
\ee 
In this case,  $F_{(\gb i)}=-\p_{(\gb i)}U$ where
\be\label{e:swimmer_potential}
U=\sum_\gs U^\gs,
\qquad\qquad
U^\gs(t,d^\gs)
=
\f{k_0}{2}\,[d^\gs- L^\gs(t) ]^2.
\ee
The dumbbell swimmer is called \emph{passive} if the stroke amplitude is zero,  $\gl=0$, and \emph{active} if $|\gl|>0$. As discussed below, the phase parameter $\gvf^\gs$ is important for the interaction between two or more dumbbells. 
\par
For the overdamped description~(\ref{e:langevin}) to remain valid, the driving must be sufficiently slow.  More precisely, we have to impose that $T_\gc \ll T_0 \ll T_\go$, where $T_\go:=2\pi/\go$ is the driving period, $T_{0}:=2\pi/ \sqrt{k_0/M_\ga}$ the oscillator period for  a sphere of mass $M_\ga$,  and  $T_\gc:=M_{\ga}/\gc_\ga$ the characteristic damping time. This restriction ensures that  the dumbbells  behave similar to shape-driven swimmers, i.e., $d^\gs\simeq L^\gs(t)$ is a useful approximation in analytical calculations.

%%%%%%%%%%%%%%%%%%%%%%%%
\subsection{Coarse-grained mesoscopic dynamics}
\label{s:meso}
%%%%%%%%%%%%%%%%%%%%%%%%

We next summarize the stroke-averaged equations of motions  for the dumbbell positions and orientations, obtained by applying the procedure described in the Appendix of Ref.~\cite{2010PuDuYe}.  Below, the resulting effective equations of motion will be compared with numerical simulations of the microscopic model equations~(\ref{e:langevin}). 
\par
Each dumbbell can be characterized by its orientation vector
\be
\tilde{\bs N}^\gs(t)=\f{\bs X_{2\gs}-\bs X_{2\gs-1}}{|\bs X_{2\gs}-\bs X_{2\gs-1}|},
\qquad
\gs=1,\ldots, S
\ee 
and a suitable position coordinate
\be
\tilde{\bs R}^\gs(t)=\gb_2 \bs X_{2\gs}+\gb_1\bs X_{2\gs-1},
\qquad
\gb_1+\gb_2=1,\qquad\gb_{1/2}>0.
\ee
For example,  the choice $\gb_1=\gb_2=1/2$ corresponds to the \emph{geometric} center~\cite{2008LaBa}
\be\label{e:geom_center}
\tilde{\bs R}_\mrm{G}^\gs(t):=\f{1}{2}\left(\bs X_{2\gs}+\bs X_{2\gs-1}\right),
\ee
Here, we shall consider $\gb_i=a_i/(a_1+a_2)$, defining the \emph{center of hydrodynamic stress} ~\cite{HappelBrenner,2009BaMa_PNAS} 
\be\label{e:hydro_center}
\tilde{\bs R}_\mrm{H}^\gs(t):=\f{a_2\bs X_{2\gs}+a_1\bs X_{2\gs-1}}{a_1+a_2}.
\ee
For very small asymmetries $a_1\approx a_2$, the geometric center $\tilde{\bs R}^\gs_\mrm{G}$ practically coincides  with the hydrodynamic center $\tilde{\bs R}^\gs_\mrm{H}$. For strongly asymmetric dumbbells with $a_1\ll a_2$ or $a_1\gg a_2$ the hydrodynamic center $\tilde{\bs R}^\gs_\mrm{H}$ presents the more appropriate choice, as it is the \lq slower\rq\space variable.
\par
The basic idea of the stroke-averaging procedure~\cite{2008LaBa,2008AlYe,2009DuZa} is to focus on the dynamics of averaged position and orientation coordinates $\bs R(t)$ and  $\bs N^\gs(t)$,  defined by
\be
\bs N^\gs(t)
:=
\f{1}{T}\int_{t-T/2}^{t+T/2}\ds\;
\tilde{\bs N}^\gs(s)
\csp
\bs R^\gs(t)
:=
\f{1}{T}\int_{t-T/2}^{t+T/2}\ds\;
\tilde{\bs R}^\gs(s).
\ee
Here $T=2\pi/\go$ denotes the period of a swimming stroke. If $\tilde{\bs N}^\gs(t)$  and   $\tilde{\bs R}^\gs(t)$ are slowly varying functions of time, one can  approximate
\be\label{e:stroke_app}
\dot{\bs N}^\gs
\simeq
\dot{\tilde{\bs N}}^\gs
\csp
\dot{\bs R}^\gs
\simeq
\dot{\tilde{\bs R}}^\gs,
\qquad
\f{1}{T}\int_{t-T/2}^{t+T/2}
\!\!\!\!\ds\; f(\tilde{\bs N}^\gs(s),\tilde{\bs R}^\gs(s))
\simeq
 f({\bs N}^\gs(t),{\bs R}^\gs(t))
\ee
for any sufficiently well-behaved function $f$. For nearly symmetric dumbbells, the approximations~(\ref{e:stroke_app}) are justified for both $\tilde{\bs R}_\mrm{G}$ and  $\tilde{\bs R}_\mrm{H}$, whereas for strongly asymmetric ones they usually only hold for the hydrodynamic center $\tilde{\bs R}_\mrm{H}$.
\par
Using Eqs.~(\ref{e:stroke_app}), one can derive  from the microscopic model equations~(\ref{e:langevin}) the corresponding deterministic stroke-averaged equations of motions~\cite{2008LaBa,2008AlYe,2009DuZa}, by assuming that: (i) The dumbbells are force-free~\footnote{If the internal forces required contract the dumbbells are central forces, then the force-constraint implies that the torque-free constraint is automatically fulfilled.} and approximately shape driven,  i.e., ~$d^\gs:=|\bs X_{2\gs}-\bs X_{2\gs-1}|\simeq L^\gs(t)$. (ii) The dumbbells are slender, i.e., sphere radii  $a_{1/2}$ and stroke amplitude $\gl$  have about the same size, but are much smaller than the dumbbell's mean length $\ell$.
(iii) The ensemble is dilute, meaning that the distance $D_\mrm{G/H}^{\gs\gr}:=|\bs D_\mrm{G/H}^{\gs\gr}|:=|\bs R_\mrm{G/H}^\gs-\bs R_\mrm{G/H}^\gr|$ between dumbbells $\gs$ and $\gr$   is much larger than $\ell$. 
\par
Adopting the simplifications (i)--(iii) and restricting to two-body interactions, one finds the following coarse-grained  equations of motion for the hydrodynamic center,
\be\label{e:eom}
\dot{R}_{\mrm{H} i}^\gs &=
&
\left(\f{a_2^2-a_1^2}{a_1^2+a_2^2}\right)\sum_{\gr\ne \gs}  I^{\gs\gr}_i  +  \sum_{\gr\ne\gs} J^{\gs\gr}_i,
\\
\dot{N}_i^\gs &=
&
\label{e:eom_a} 
- (\gd_{ik}- N^\gs_iN^\gs_k)\;
\left\{
\left(\f{a_2-a_1}{a_1+a_2}\right)\sum_{\gr\ne \gs} K^{\gs\gr}_k +
\sum_{\gr\ne\gs} L^{\gs\gr}_k\right\},
\ee
where the stroke-averaged hydrodynamic interaction terms to leading order in $\gl/\ell$ are given by
\be
I^{\gs\gr}_i
&=&\nonumber
\Gl\go\,\sin(\gvf^\gs-\gvf^\gr)\;
\f{9} {64} 
\left(\f{\gl}{\ell}\right)^2
\left(\f{\Gl}{\ell}\right)
\left(\f{\ell}{|\bs D_\mrm{H}^{\gs\gr}|}\right)^3\;\times
\\&&
N^\gs_i  
\bigl(
1-3r^2 -3s^2-6qsr+
15s^2r^2 
\bigr),
\\
J^{\gs \gr}_i
&=&\nonumber
\Gl\go\,
\sin(\gvf^\gs-\gvf^\gr)\;\f{9}{64}
\left(\f{\gl}{\ell}\right)^2
\left(\f{\ell}{|\bs D_\mrm{H}^{\gs\gr}|}\right)^4\times
\\&&
\bigl\{
N^\gs_{i}
(2s+4qr-10 sr^2) +
\hat{D}_{\mrm{H}i}^{\gs\gr}(1+ 2 q^2-5 s^2-5r^2- 20qsr  + 35 s^2 r^2)
\bigr\},
\\
%\ee
%and
%\be
K^{\gs\gr}_k
&=&\nonumber
\go\sin(\gvf^\gs - \gvf^\gr)\;\f{9}{32}
\left(\f{\gl}{\ell}\right)^2
\left(\f{\Gl}{\ell}\right)
\left(\f{\ell}{|\bs D_\mrm{H}^{\gs\gr}|}\right)^4\times
\\&&
\hat{D}_{\mrm{H}k}^{\gs\gr}(1+ 2 q^2-5 s^2-5r^2- 20qsr  + 35 s^2 r^2),
\\
L^{\gs\gr}_k 
&=&\nonumber
\go\sin(\gvf^\gs-\gvf^\gr)\;\f{15}{64}
\left(\f{\gl}{\ell}\right)^2
\left(\f{\Gl}{\ell}\right)
\left(\f{\ell}{|\bs D_\mrm{H}^{\gs\gr}|}\right)^5\times
\\&&
\hat{D}_{\mrm{H} k}^{\gs\gr}
 \bigl( 3s +6rq+ 6sq^2 -7s^3 -  21sr^2-
% \quad\notag
%\\&&\qquad\;
42qs^2r + 63 s^3r^2
\bigr).
\label{e:eom_last}
\ee
Here, 
\be
\Gl:=2a_1a_2/(a_1+a_2)
\ee 
denotes the harmonic mean of the sphere radii, the unit vector $\hat{\bs D}_\mrm{H}^{\gs\gr}:={\bs D_\mrm{H}^{\gs\gr}}/{|\bs D_\mrm{H}^{\gs\gr}|}$ gives the orientation of the distance vector $\bs D_\mrm{H}^{\gs\gr}=\bs R_\mrm{H}^\gs-\bs R_\mrm{H}^\gr$, and $s,r,q$ abbreviate the projections
\be
s:=\hat D_{\mrm{H} j}^{\gs\gr}N_j^\gs,
\qquad
r:=\hat D_{\mrm{H} j}^{\gs\gr} N_j^\gr,
\qquad
q:=N^\gs_jN^\gr_j.
\quad
\ee
One readily observes two prominent features: Firstly, the stroke-averaged interactions terms $I,J,K,L$ vanish if the phases $\gvf^\gs$ and $\gvf^\gr$ differ by multiples of $\pi$. Secondly, in the dilute limit  $|\bs D_\mrm{H}^{\gs\gr}|\gg \ell$ the leading contribution to $\dot{\bs R}_\mrm{H}^\gs$ is given by the $I$-terms, which decay as $|\bs D_\mrm{H}^{\gs\gr}|^{-3}$~\cite{2008LaBa}. By contrast, for symmetric dumbbells  with $a_1=a_2$ the first term in Eq.~(\ref{e:eom_a}) is absent and one recovers the equations recently derived in Ref.~\cite{2008AlYe}; hence, in this case, the effective interaction force decays asymptotically as~$|\bs D_\mrm{H}^{\gs\gr}|^{-4}$.  
Similarly, asymmetric dumbbells experience changes in rotation $\dot{N}^\gs$ which decay as $|\bs D_\mrm{H}^{\gs\gr}|^{-4}$, whereas the effective rotation interaction for symmetric dumbbells decays as $|\bs D_\mrm{H}^{\gs\gr}|^{-5}$.

%%%%%%%%%%%%%%%%%%%%%%%%
\section{Numerical tests}
\label{s:numerics}
%%%%%%%%%%%%%%%%%%%%%%%%

We verify the validity of the stroke-averaged equations of motion~(\ref{e:eom}) by comparing with time-resolved 
simulations of the microscopic dynamics~(\ref{e:langevin}). To this end, we numerically simulate both microscopic and coarse-grained equations of motions using a CUDA algorithm that is described in detail in Ref.~\cite{2010PuDuYe}. 
We first compare our results to those of Lauga and Bartolo~\cite{2008LaBa} for the one-dimensional case. Our findings agree with theirs for most setups, they differ for one particular configuration (see discussion below). 
We conclude our numerical analysis by testing the coarse-grained equations of motions for various  three-dimensional dumbbell configurations.
\par
Generally, when comparing symmetric and asymmetric dumbbell configurations, one should keep the quantity  $\Gl=2a_1a_2/(a_1+a_2)$ constant, in order to keep the \lq symmetric\rq\space $J$-contribution in Eq.~(\ref{e:eom_a}) fixed.

%%%%%%%%%%%%%%%%%%%%%%%%
\subsection{One-dimensional case}
\label{s:1d}
\begin{figure}[t!]
\centering
\resizebox{0.47\columnwidth}{!}{\includegraphics{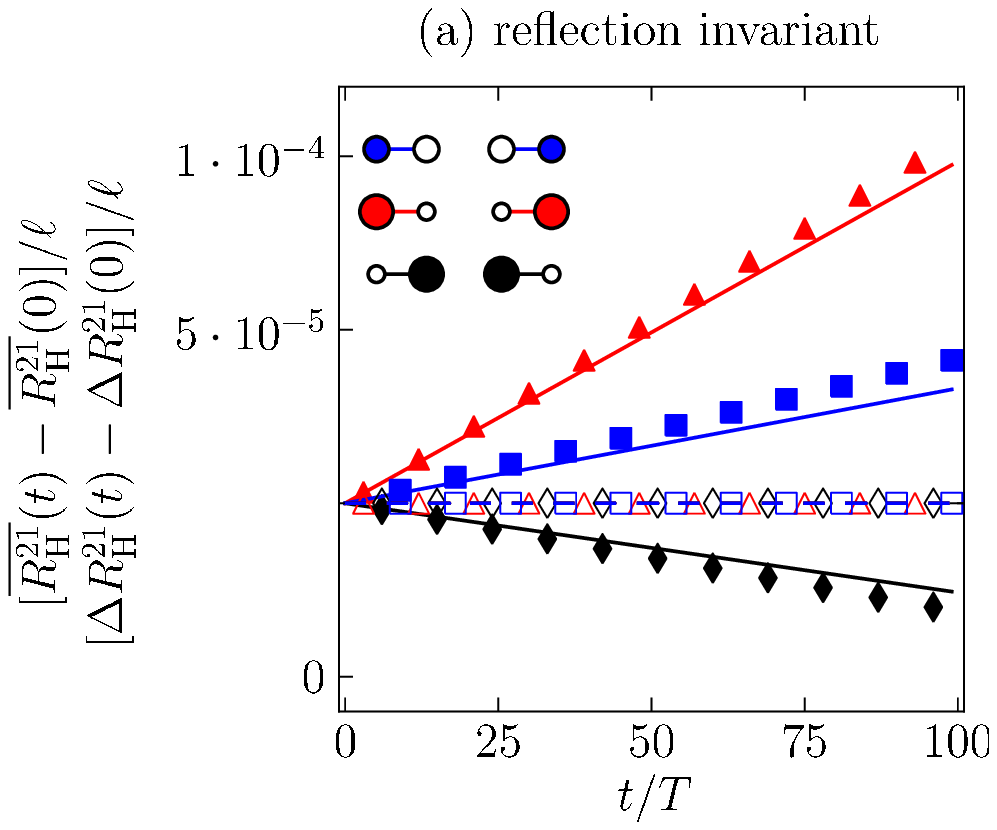}}
\hspace{0.2cm}
\resizebox{0.47\columnwidth}{!}{\includegraphics{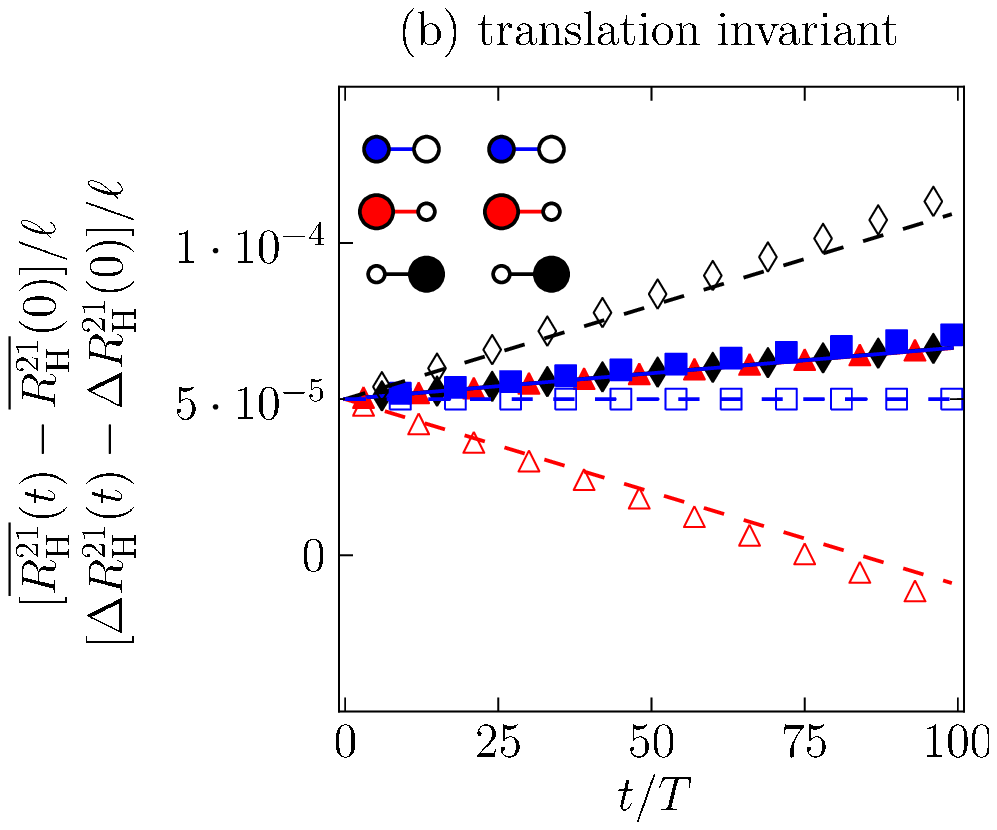}}
\caption{Depending on the symmetry of their initial configuration,  aligned dumbbell pairs exhibit qualitatively different collective motions.  In both diagrams, lines were obtained by numerical integration of the stroke-averaged equations~(\ref{e:stroke_averaged}), whereas symbols show the simulation results for the microscopic spring-based dumbbell model~(\ref{e:langevin}).  Solid lines and filled symbols depict the mean displacement  \mbox{$\ovl{R^{21}_\mrm{H}}(t)-\ovl{R^{21}_\mrm{H}}(0)=\{[R^2_\mrm{H}(t)+R^1_\mrm{H}(t)]-[R^2_\mrm{H}(0)+R^1_\mrm{H}(0)]\}/2$} of the hydrodynamic centres. Dashed lines and unfilled symbols indicate the relative separation  \mbox{$\Gd R_\mrm{H}^{21}(t)-\Gd R_\mrm{H}^{21}(0)=[R^2_\mrm{H}(t)-R^1_\mrm{H}(t)]-[R^2_\mrm{H}(0)-R^1_\mrm{H}(0)]$}.  (a)~For mirror symmetric configurations, the mean distance between the dumbbells remains constant and the dumbbell pair  can move in either direction. In particular, asymmetry can enhance the collective speed, see triangles.  (b) For translation invariant configurations, asymmetry does not affect the collective pair velocity, but depending on the initial orientation the dumbbells either approach each other (red triangles) or move away from each other (black diamonds).   
Simulation parameters are comparable to those of Lauga and Bartolo~\cite{2008LaBa}: Initial separation $\Gd R_\mrm{H}^{21}(0)=R^2_\mrm{H}(0)-R^1_\mrm{H}(0)=10\ell$, mean dumbbell length $\ell=5 \mu \met$, driving frequency $\go=500\sec^{-1}$ (time is given in units of the stroke period $T=2\pi/\go$),  stroke amplitude $\gl=0.1\ell$ , $\Gl=2a_1a_2/(a_1+a_2)=0.15\ell$ with $a_2=a_1/2$ for asymmetric dumbbells, phase difference $\gvf^2- \gvf^1=\pi/2$.  For the microscopic model: spring constants $k_0=0.001\kg/\sec^2$, viscosity $\mu=10^{-3}$ kg/(ms), particle mass density  $\gvr=10^{3}$~kg/m$^3$; simulation time step $\Gd t\approx10^{-4}T$. We note that the quantitative difference between symbols and lines decreases when choosing smaller ratios $(a_\ga/\ell)$, cf. discussion in Sec. 4.1 of Ref.~\cite{2010PuDuYe}.}
\label{fig01}
\end{figure}

%%%%%%%%%%%%%%%%%%%%%%%%
We first consider aligned dumbbell pairs (see Fig.~\ref{fig01}) as studied by  Lauga and Bartolo~\cite{2008LaBa}.  
Aligned dumbbells do not change their orientation and Eqs.~(\ref{e:eom_a}) reduce to
\be
\dot{R}_\mrm{H}^\gs
&=& \nonumber
\f{9}{16} \sum_{\gr\ne \gs}
\Gl\go \sin(\gvf^\gs-\gvf^\gr)\;
\left(\f{\gl}{\ell}\right)^2 \times
\\&&      
\;\biggl\{
\biggl(\f{a_2^2-a_1^2}{a_2^2+a_1^2}\biggr)\biggl(\f{\Gl}{\ell}\biggr)\biggl(\f{\ell}{| D_\mrm{H}^{\gs\gr}|}\biggr)^3 N^\gs+ 
\biggl(\f{\ell}{|D_\mrm{H}^{\gs\gr}|}\biggr)^4 \hat{D}_\mrm{H}^{\gs\gr}
\biggr\},
\quad
\label{e:stroke_averaged}
\ee
where $\dot{R}_\mrm{H}^\gs$ denotes the coordinates along the common axis. 
The lines in Figs.~\ref{fig01} (a) and (b) represent the dynamics of aligned dumbbell pairs as predicted by Eq.~(\ref{e:stroke_averaged}), and symbols indicate the results of  corresponding microscopic model simulation. 
Following Lauga and Bartolo~\cite{2008LaBa}, we quantify collective motion of the dumbbell pairs in terms of their 
mean collective displacement (solid lines/filled symbols in Fig~\ref{fig01}),
\be
\ovl{R^{21}_\mrm{H}}(t)
=
\f{1}{2}[R^2_\mrm{H}(t)+R^1_\mrm{H}(t)],
\quad
\ee
and their mean relative distance  (dashed lines/unfilled symbols in Fig~\ref{fig01}),
\be
\Gd R_\mrm{H}^{21}(t)
=
R^2_\mrm{H}(t)-R^1_\mrm{H}(t).
\quad
\ee
The quantity $\ovl{R^{21}_\mrm{H}}(t)$ characterizes the net motion of the dumbbell pair, whereas $\Gd R_\mrm{H}^{21}(t)$ indicates whether the dumbbells move towards or away from each other.
%%%%%%%%%%%%%%%%%%%%%%%%%%%%%%%%%%%%%%%%%%%
%%%%%%%%%%%%%%%%%%%%%%%%%%%%%%%%%%%%%%%%%%
\par
As evident from the two diagrams in Fig.~\ref{fig01}, we find qualitatively different behavior depending on the symmetry of the initial dumbbell pair configuration.   Figure~\ref{fig01}~(a) summarizes results for reflection symmetric initial conditions. In this case,   the pair can move in either direction, depending on whether the smaller spheres point towards or away from each other, but the mean distance between the two dumbbells remains constant. By contrast, the translation invariant configurations in Fig.~\ref{fig01}~(b) always move in the same direction with nearly identical speeds for symmetric and asymmetric dumbbell pairs. We note that our results for the translation invariant setups essentially agree with those obtained by Lauga and Bartolo~\cite{2008LaBa}. However,  we find a different behavior for mirror symmetric configurations, cf. Fig. 2~(a) of their paper~\cite{2008LaBa}. In particular, our results show that under suitable conditions collective motions can be significantly enhanced by asymmetry, see red triangles in Fig~\ref{fig01}~(a).  The fact that asymmetry can increase effective hydrodynamic interactions could be of relevance for the construction of more efficient micropumps~\cite{2009LeEtAl}.
\par
\begin{figure}[t!]
\centering
\resizebox{0.47\columnwidth}{!}{\includegraphics{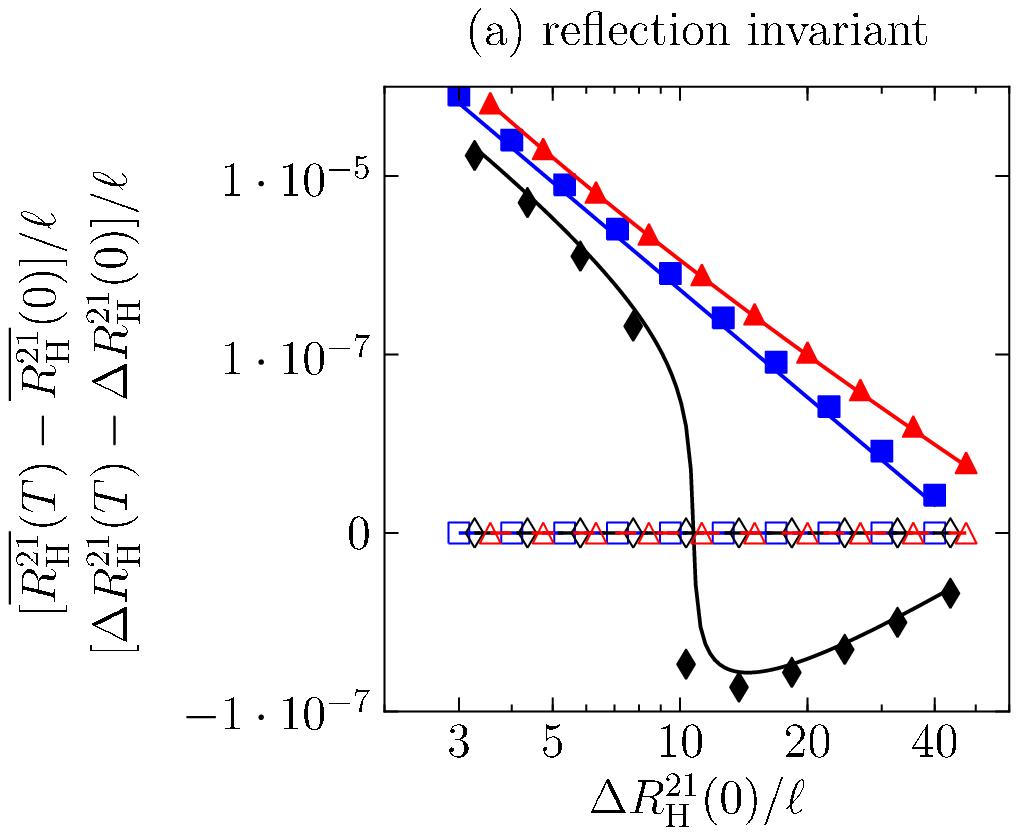}}
\resizebox{0.47\columnwidth}{!}{\includegraphics{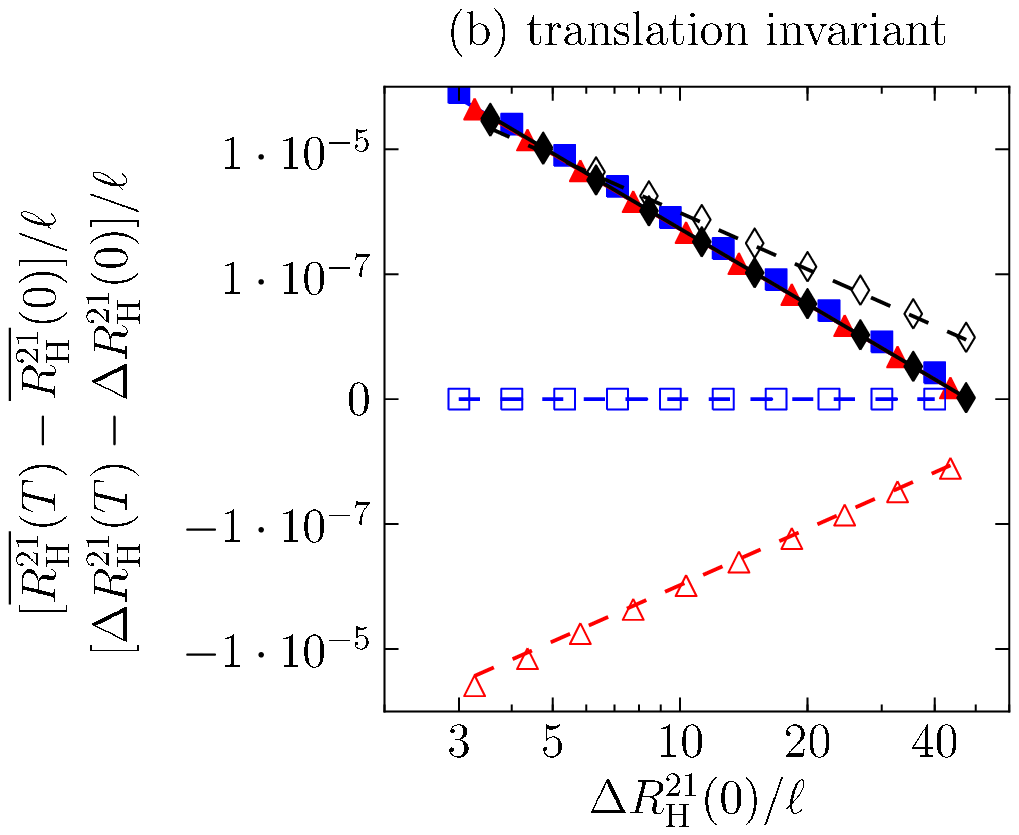}}
\caption{Distance dependance of the collective motion and separation for aligned dumbbell pairs.  Line styles and symbols correspond to the same configurations and simulation parameters as used in Fig.~\ref{fig01}. For a reflection symmetric configuration such that the small spheres point outwards, see  black diamonds in diagram (a), the collective velocity reverses its sign at a distance of approximately~$11\ell$. Qualitatively, this change is correctly reflected by the stroke-averaged equation~(\ref{e:eom}), see solid black line in diagram (a). Remarkably, the stroke-averaged equations describe the microscopic dumbbell dynamics reasonably well down to distances of a few body lengths.
}
\label{fig02}
\end{figure}

Furthermore, Fig.~\ref{fig02} shows how both the collective displacement and the relative separation per period vary with the distance between the dumbbells. As in Fig.~\ref{fig01}, lines indicate the prediction based on the stroke-averaged equations~(\ref{e:eom}), whereas symbols indicate the simulation results for the related spring-based model (symbols, colors, and line-styles refer to the same configurations/model parameters as in Fig.~\ref{fig01}, respectively). Remarkably, the   dynamics of the microscopic model is very well described by the averaged equations~(\ref{e:eom}) down to distances of a few body lengths. This is a bit surprising given that the stroke-averaged equations are based on a far-field expansion. In particular, as also correctly predicted by the stroke-averaged equations~(\ref{e:eom}), for the mirror symmetric configuration with small spheres pointing outwards  the pair velocity reverses its sign at a distance of approximately~$11\ell$; see black diamonds in Fig.~\ref{fig02}~(a).
\par

%%%%%%%%%%%%%%%%%%%%%%%%%%%%%%%%%%%%%%%%%%%
%%%%%%%%%%%%%%%%%%%%%%%%%%%%%%%%%%%%%%%%%%

%%%%%%%%%%%%%%%%%%%%%%%%
\subsection{Three-dimensional case}
\label{s:3d}
%%%%%%%%%%%%%%%%%%%%%%%%

\begin{figure}[t!]
\centering
\resizebox{0.47\columnwidth}{!}{\includegraphics{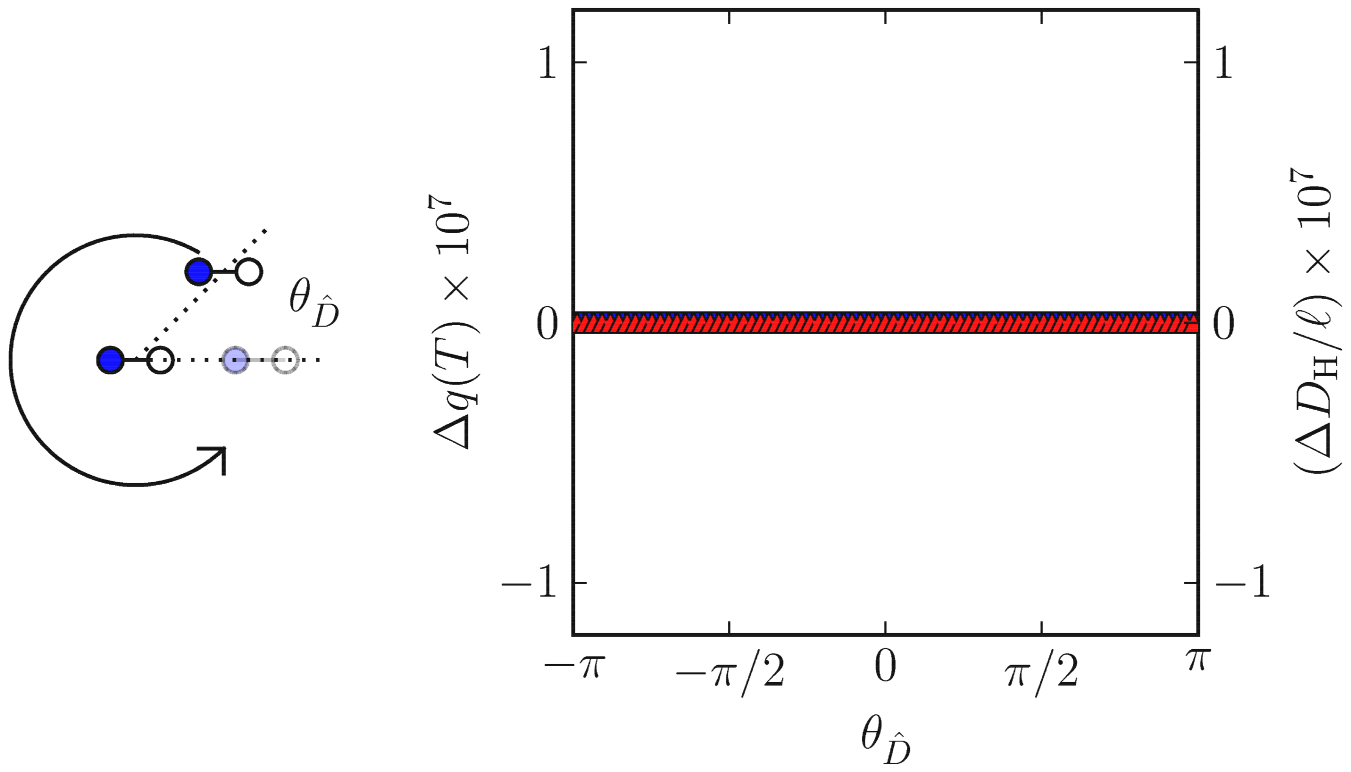}}
\resizebox{0.47\columnwidth}{!}{\includegraphics{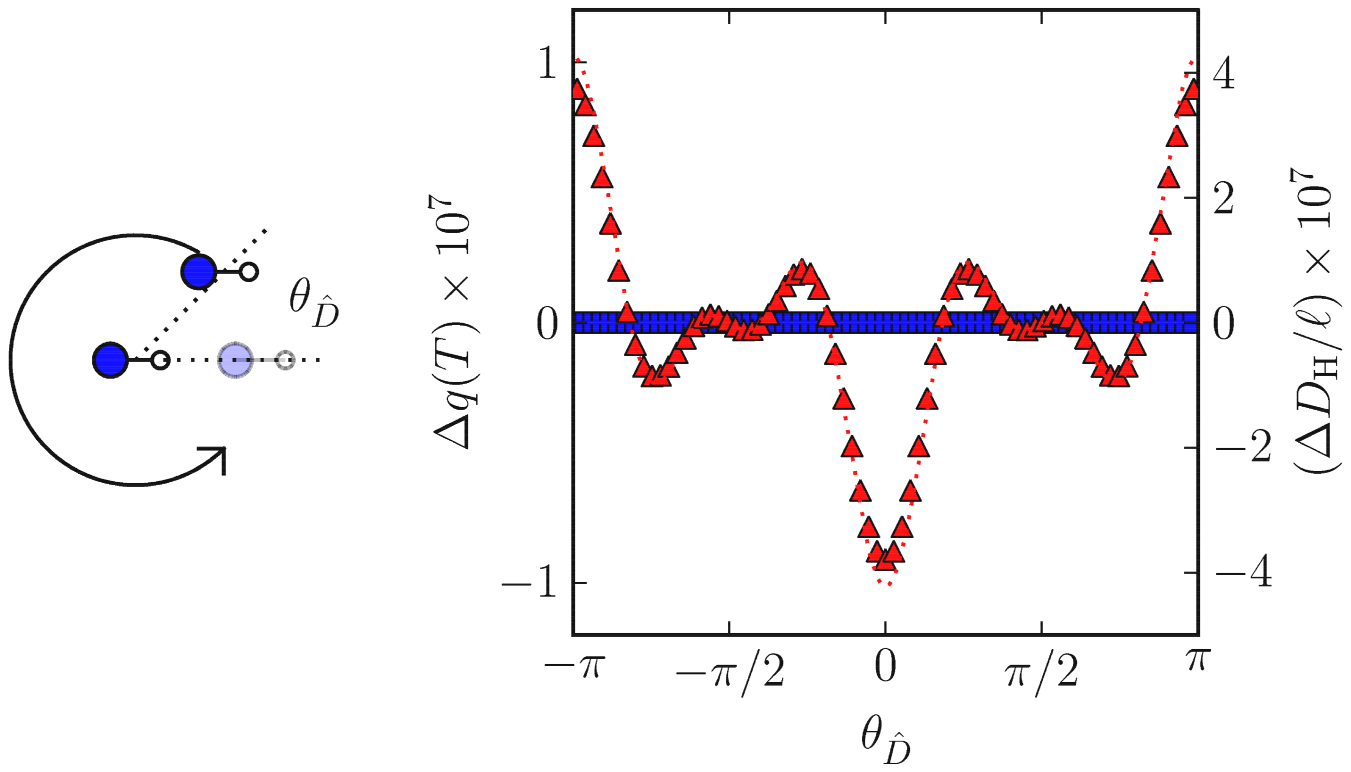}}\\
\resizebox{0.47\columnwidth}{!}{\includegraphics{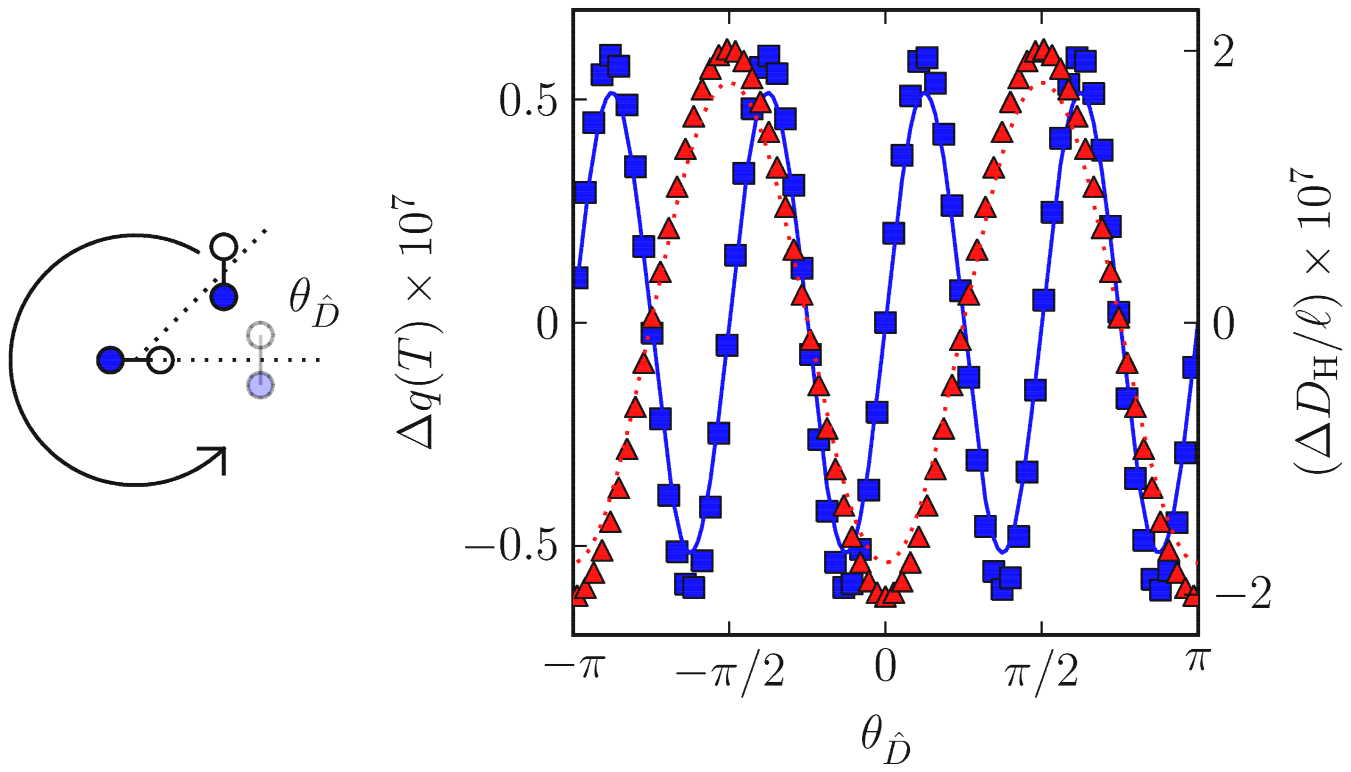}}
\resizebox{0.47\columnwidth}{!}{\includegraphics{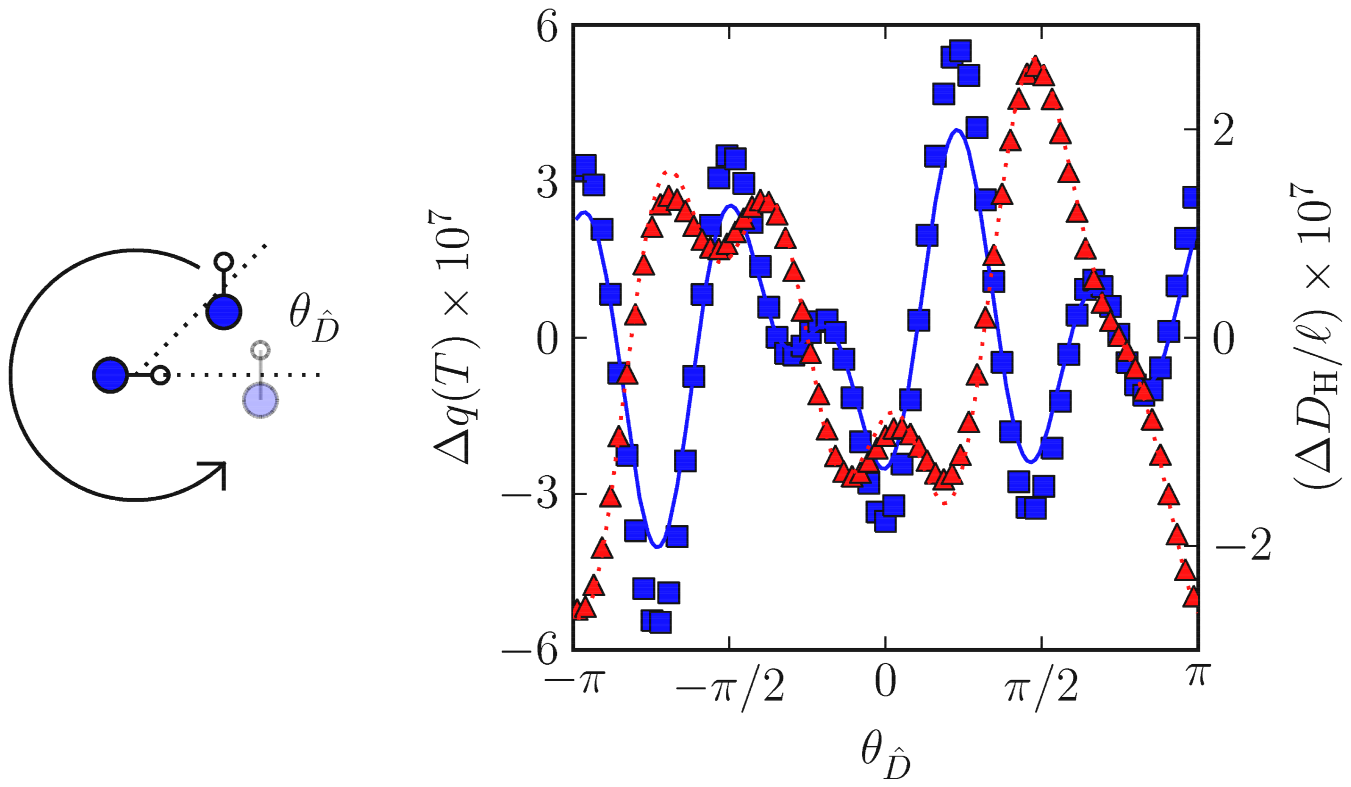}}\\
\resizebox{0.47\columnwidth}{!}{\includegraphics{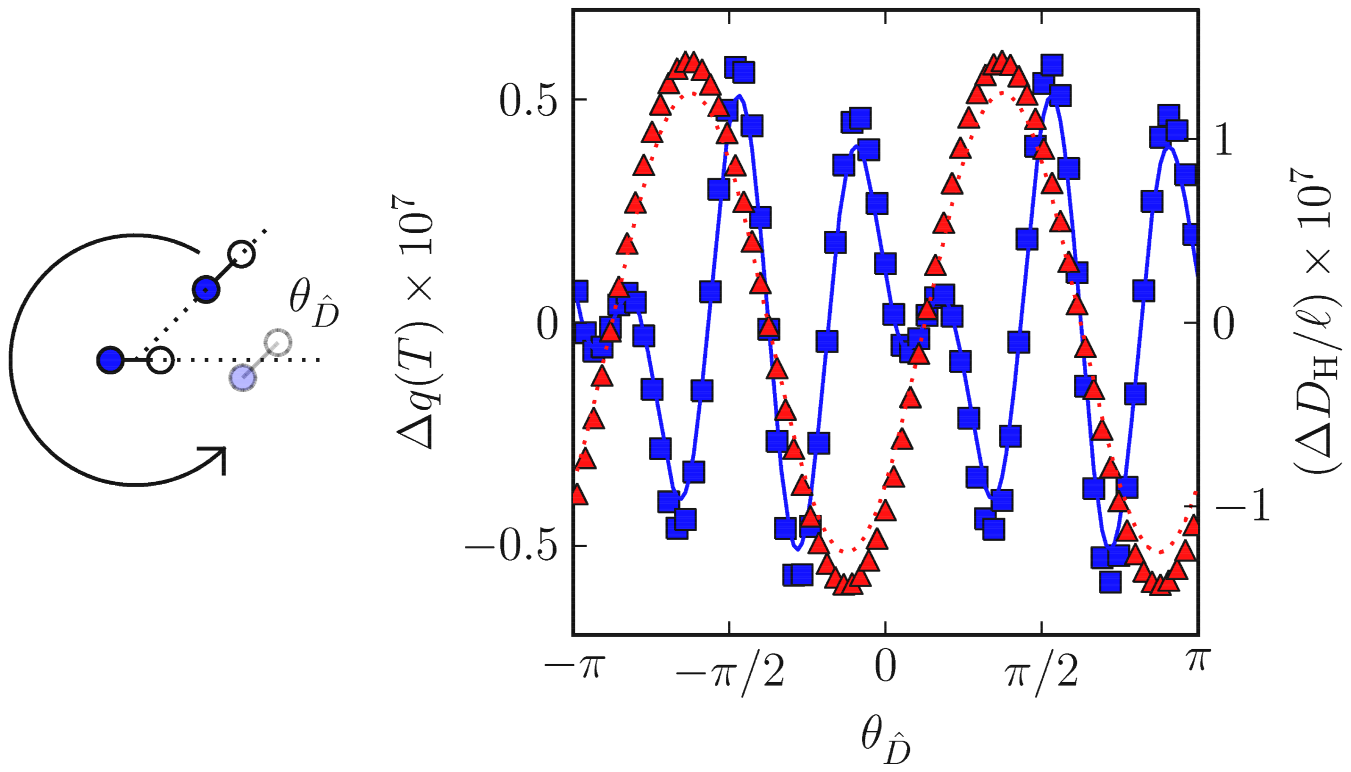}}
\resizebox{0.47\columnwidth}{!}{\includegraphics{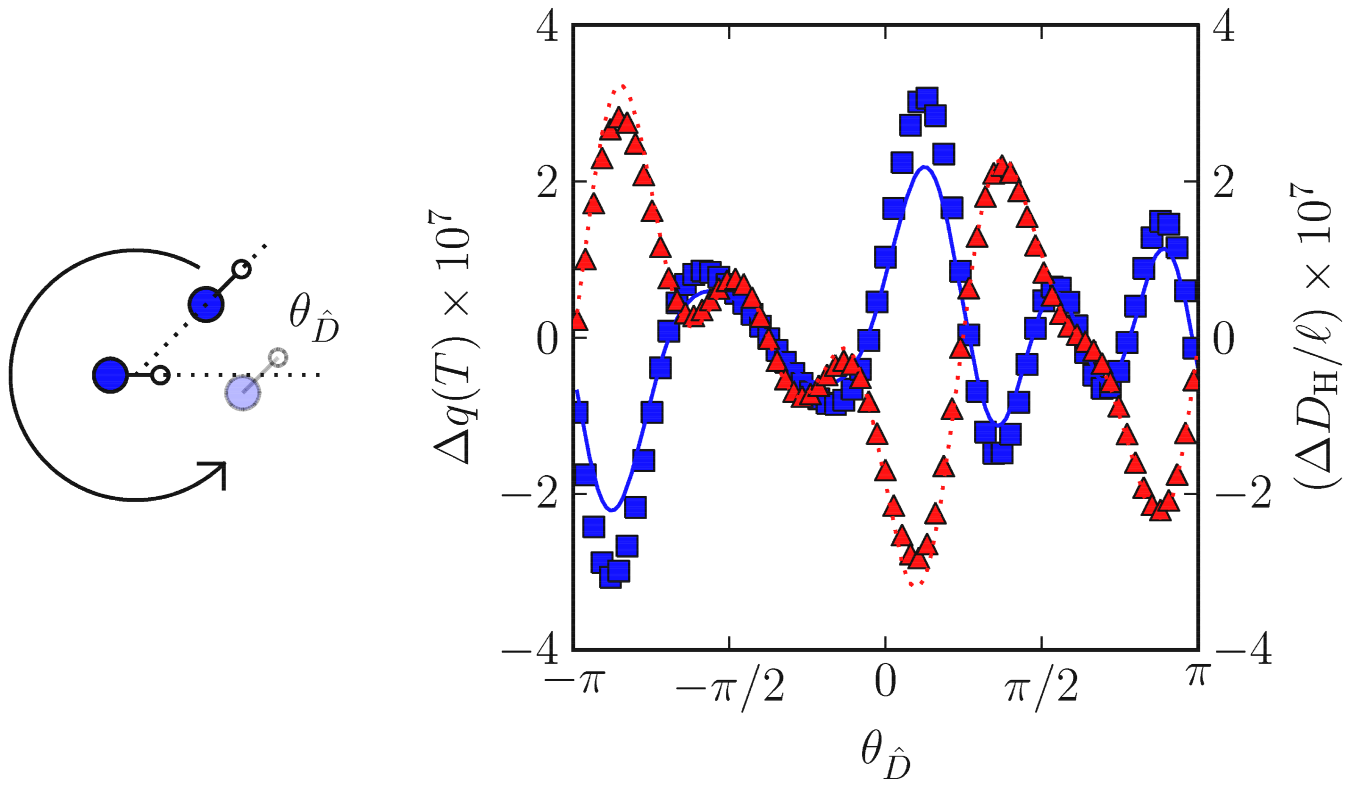}}
\caption{
In-plane rotation of the initial swimmer positions. Symbols are obtained by numerical simulations of the microscopic equations of motions and lines indicate the results of the stroke-averaged dynamics. Blue solid lines/blue squares denote the change of the relative orientation $\Gd q$ over a stroke period. Red dotted lines/red triangles indicate the change of the relative distance $\Gd D_\mrm{H}$ over a stroke period.  Swimmer parameters are identical to those used in the  1d tests except for $\Gl=2 a_1 a_2 / (a_1 + a_2) = 0.1\ell$.
}
\label{fig_in_plane}
\end{figure}

\begin{figure}[t!]
\centering
\resizebox{0.47\columnwidth}{!}{\includegraphics{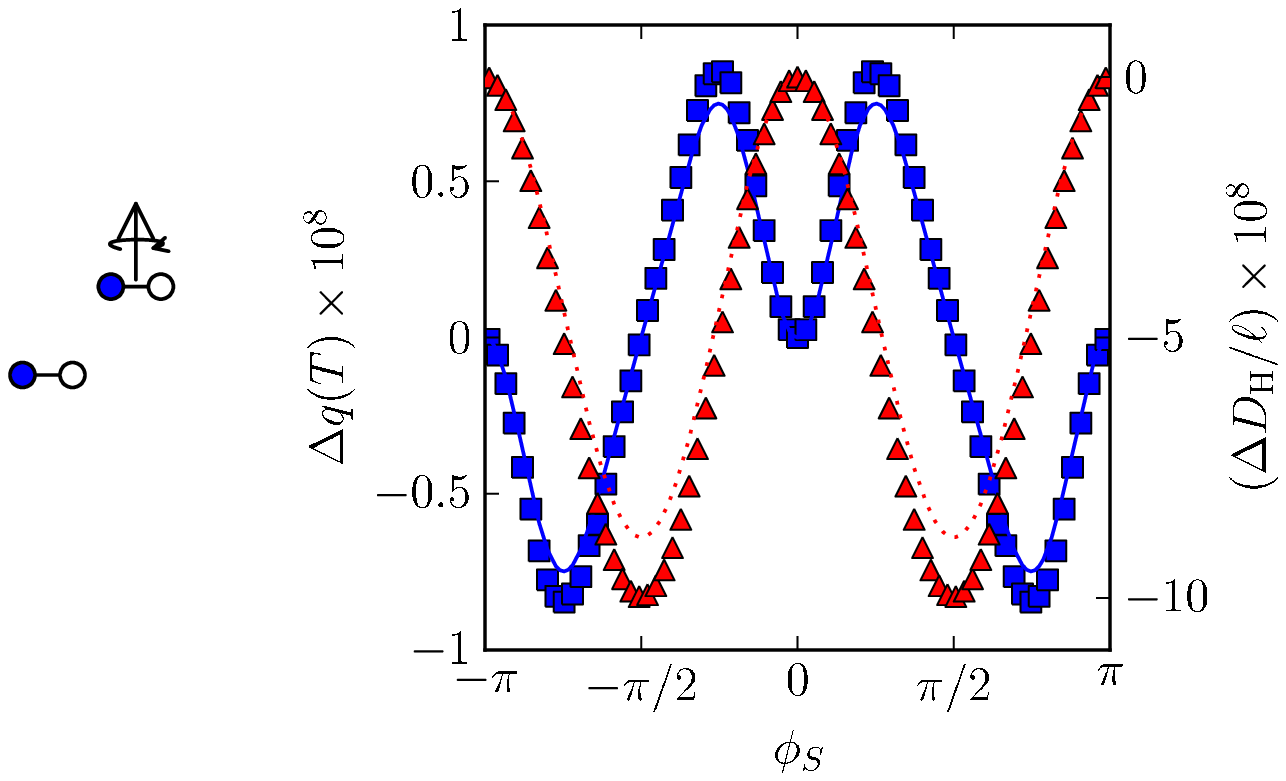}}
\resizebox{0.47\columnwidth}{!}{\includegraphics{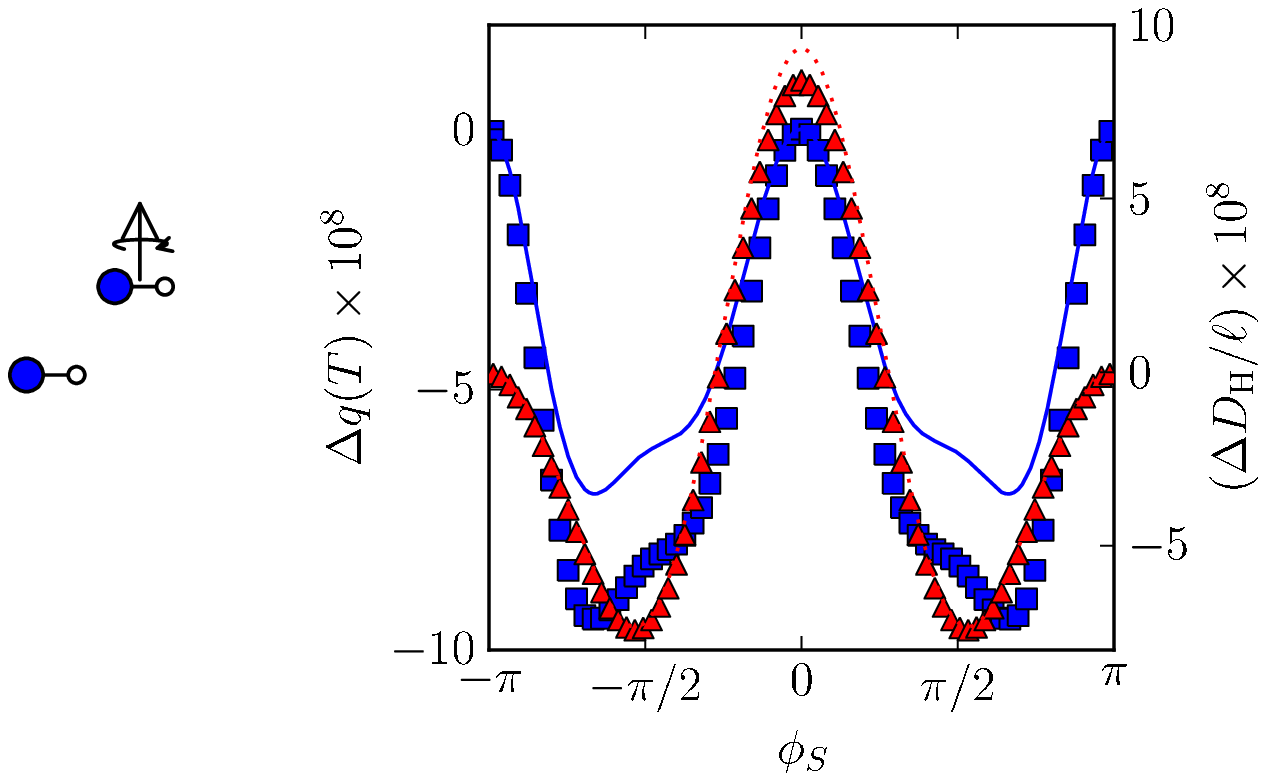}}\\
\resizebox{0.47\columnwidth}{!}{\includegraphics{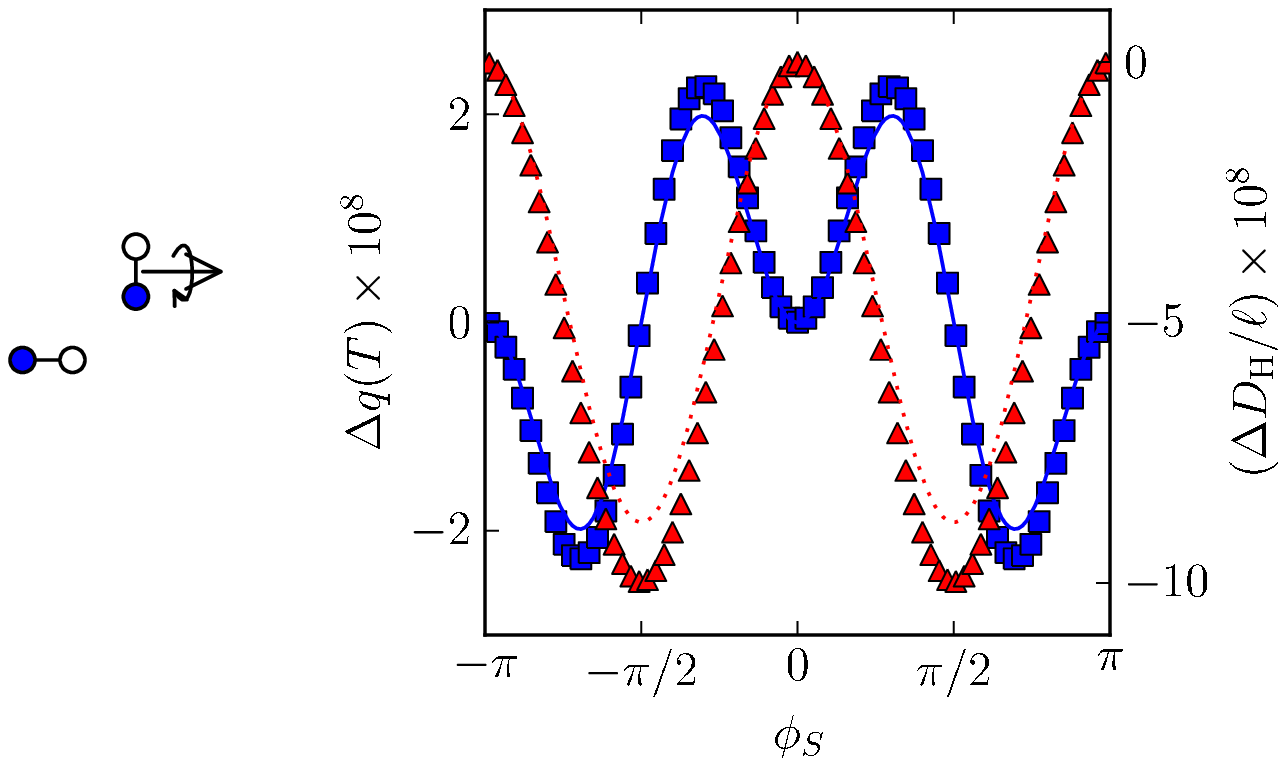}}
\resizebox{0.47\columnwidth}{!}{\includegraphics{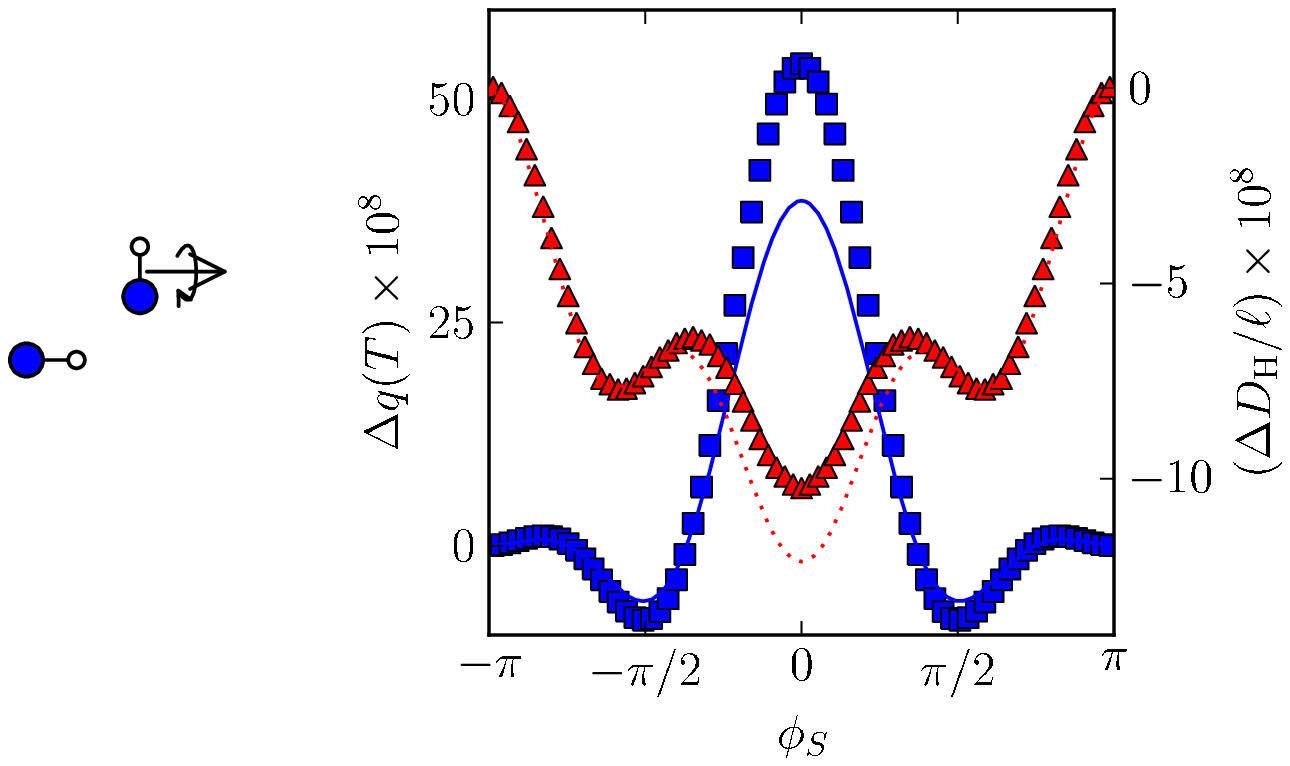}}
\caption{
Out-of-plane rotation of the initial orientation of the second swimmer by an angle $\phi_{s}$ around the indicated axis.
Symbols are obtained by numerical simulations of the microscopic equations of motions and lines indicate the results of the stroke-averaged dynamics. Blue solid lines/blue squares denote the change of the relative orientation $\Gd q$ over a stroke period. Red dotted lines/red triangles indicate the change of the relative distance $\Gd D_\mrm{H}$ over a stroke period.  Swimmer parameters are identical to those used in the  1d tests except for $\Gl=2 a_1 a_2 / (a_1 + a_2) = 0.1\ell$.}
\label{fig_out_of_plane}
\end{figure}

In two- and three-dimensional systems, hydrodynamic interactions can not only lead to collective translational but may also induce orientational changes. To illustrate both effects and to verify the validity of the angular parts in Eqs.~(\ref{e:eom})-(\ref{e:eom_last}), we consider various different initial configurations as sketched next to the diagrams in Fig.~\ref{fig_in_plane} and~\ref{fig_out_of_plane}. In each of the diagrams, blue symbols/squares indicate the change of the relative orientation of the two dumbbells, quantified through the change of the projection 
\be
\Gd q(T):=q(0)-q(T)
\csp\qquad
q(t):=N_i^\gs(t)N_i^\gr(t),
\ee
where $\bs N^\gs(t)$ and $\bs N^\gr(t)$ denote the orientation of the dumbbells $\gs$ and $\gr$ at time $t$, respectively. It is important to note that, for  symmetric dumbbells with $a_1=a_2$, the orientation vector is \emph{not} uniquely defined (in this case, the coarse-grained equations of motions are invariant under the transformation $\bs N\mapsto -\bs N$). For asymmetric dumbbells, however, the orientation can be uniquely characterized by means of the different sphere radii. In our plots, we fix the orientation as pointing from the filled to the unfilled sphere in all cases.
\par
Furthermore, the red lines/triangles show the change in the relative separation of the two dumbbells after one period,
\be
\Gd D_\mrm{H}(T):=\left[D_\mrm{H}(0)-D_\mrm{H}(T)\right]/\ell
\csp
D_\mrm{H}(t):=|\bs R^\gs_\mrm{H}(t)-\bs R^\gr_\mrm{H}(t)|/\ell.
\ee
For $\Gd D_\mrm{H}(T)>0$ the dumbbells approach each other; for $\Gd D_\mrm{H}(T)<0$ they move away from each other.
\par
To illustrate the effects of a change in symmetry,  we plot the results for symmetric and asymmetric dumbbell pairs next to each other. As one may readily observe, asymmetry does strongly affect both the translational and orientational motions of the dumbbells. This suggests that geometry can play an important role for the emergence of orientational (dis)order in suspensions of hydrodynamically interacting organisms. Generally, Figs.~\ref{fig_in_plane} and~\ref{fig_out_of_plane} confirm the coarse-grained equations of motions~(\ref{e:eom})-(\ref{e:eom_last}) are able to quantitatively reproduce the characteristic features of the microscopic dynamics.

%%%%%%%%%%%%%%%%%%%%%%%%%%%%%%%%%%%%%%%%%%%
%%%%%%%%%%%%%%%%%%%%%%%%%%%%%%%%%%%%%%%%%%

%%%%%%%%%%%%%%%%%%%%%%%%%%%%%%%%%%%%%%%%%%%
%%%%%%%%%%%%%%%%%%%%%%%%%%%%%%%%%%%%%%%%%%

%%%%%%%%%%%%%%%%%%%%%%%%
\section{Summary}
\label{s:summary}
%%%%%%%%%%%%%%%%%%%%%%%%

Active dumbbell suspensions present a useful model for studying collective motion and orientational  ordering due to hydrodynamic interactions at low Reynolds number. In this paper, we have discussed the coarse-grained equations of motion for the effective hydrodynamic interaction between actively oscillating, asymmetric dumbbell pairs. Our analysis shows that the stroke-averaged equations~(\ref{e:eom})-(\ref{e:eom_last}) are able to capture the main features of the time-resolved, microscopic dynamics at moderate-to-low densities (intermediate-to-large distances). Thus, equations of the type~(\ref{e:eom})-(\ref{e:eom_last})  provide a convenient mesoscopic description which, for example,  can be used as a starting point for the derivation of coarse-grained macroscopic field theories~\cite{2008BaMa}. 
\par
The coarse-grained equations of motions discussed above can be readily implemented in GPU-based  CUDA simulations analogous to those described in Ref.~\cite{2010PuDuYe}.  The good agreement between the averaged dynamics and the microscopic model simulation confirms that our CUDA algorithm~\cite{2010PuDuYe} works correctly even at relatively low densities, when hydrodynamics interactions effects are relatively weak and algorithms may become prone to numerical instabilities. With regard to the future, we hope that the combination of GPU-based simulation techniques and systematic coarse-graining will enable us to quantitatively compare particle simulations with continuum field theories. The asymmetric dumbbell  model considered here appears to be particularly promising in this context as it allows one to investigate how microscopic symmetry breaking affects macroscopic behavior.

\paragraph*{Acknowledgements}
~J.D. would like to thank Lutz Schimansky-Geier for many stimulating discussions  over the past years.
This work was supported by the ONR, USA (J.D.). V.P. acknowledges
support from the United States Air Force Institute of Technology.
The views expressed in this paper are those of the authors and
do not reflect the official policy or position of the United States
Air Force, Department of Defense, or the US Government.

%\begin{table}
%\caption{Please write your table caption here.}
%\begin{tabular}{lll}
%\hline\noalign{\smallskip}
%\end{tabular}
%\end{table}
%%
%\begin{thebibliography}{}
%Author, Journal \textbf{Volume}, (year) page numbers
%Author, \textit{Book title} (Publisher, place year) page numbers
% etc
%\end{thebibliography}

\bibliographystyle{epj}
%\bibliography{S_Field,StochResonance,S_Cilia,S_Experiments,S_Hydro,S_Julia,S_Particle,S_Reviews,S_BrownianSwimmer,S_Polymer_Hydro,Journals,Hanggi,BrownianMotors,GPU_Examples,S_Dumbbell,Proceedings,Schimansky}

\end{document}